# *An Effective Medium Approach to Electron Waves: Graphene Superlattices*

*Mário G. Silveirinha*[(1, 2) *] *and Nader Engheta*[(1)]

(1) University of Pennsylvania, Department of Electrical and Systems Engineering, Philadelphia, PA, U.S.A., engheta@ee.upenn.edu
(2) University of Coimbra, Department of Electrical Engineering – Instituto de Telecomunicações, Portugal, mario.silveirinha@co.it.pt

**Abstract**

We develop an effective medium approach to characterize the propagation of matter waves in periodic structures, such as graphene or semiconductor superlattices. It is proven that the time evolution of the states that are not more localized in space than the characteristic period of the structure can be described exactly through an effective Hamiltonian, and that the electronic band structure of the system can be exactly determined from the effective Hamiltonian. As an illustration of the application of the method, we characterize the mesoscopic response of graphene superlattices. It is shown that these structures may be described using simply two effective parameters: a *dispersive potential*, and an *anisotropy tensor* that characterizes the pseudospin. Our model predicts that a graphene superlattice characterized by an indefinite anisotropy tensor – such that the eigenvalues of the tensor have opposite signs – may permit the perfect tunneling of all the stationary states with a specific value of the energy when it is paired with a dual graphene superlattice with positive definite anisotropy tensor.

---

[*] To whom correspondence should be addressed: E-mail: mario.silveirinha@co.it.pt

# I. Introduction

Effective medium semi-empirical theories, such as the $k \cdot p$ method [1] or Bastard's envelope function approximation [2], have become invaluable tools to characterize the electronic properties of bulk semiconductors and related heterostructures, and associated devices. Such treatments are made possible by the fact that the most relevant physical phenomena in semiconductors are determined by the form of the electronic structure in the vicinity of some high symmetry points in the momentum space. The $k \cdot p$ methods are based on perturbation theory [3], and enable the calculation of the band structure and effective masses of different types of semiconductors (e.g. with zincblende structure).

On the other hand, stimulated by the development of electromagnetic metamaterials [4]-[6], recently there has been an intense research of methods that permit characterizing periodic structures from an effective medium perspective, typically through the introduction of an effective permittivity and an effective permeability [7]-[13]. It is thus natural to wonder if some of these ideas and methods can be extended to the characterization of matter waves, in the context of the Schrödinger equation.

The objective of this work is precisely, by generalizing our previous studies in the context of electromagnetic metamaterials [12, 13], to bridge two fields and to develop from first principles a systematic approach that enables the computation of an effective Hamiltonian that describes within some approximations the time evolution of a quantum system, and that reduces drastically the complexity of the problem. Our analysis neglects electron-electron interactions, and thus many body effects. In some cases, these effects may be modeled by an effective potential.

We apply the developed theory to the case of one-dimensional graphene superlattices. It is demonstrated that the low-energy physics in these structures can be described simply in terms of an energy dependent effective potential and an anisotropy tensor that characterizes the pseudospin. Based on this effective medium model, we predict a novel perfect tunneling effect in graphene superlattices, showing that electron waves with a specific energy can be perfectly tunneled through a nanomaterial with specific properties.

## II. Effective Medium Approach

The starting point of our analysis is the one-body Schrödinger equation for an electron in some periodic structure (e.g. a crystalline material or a superlattice),

$$\hat{H}|\psi\rangle = i\hbar \frac{\partial}{\partial t}|\psi\rangle. \tag{1}$$

Let us suppose that the initial state is described by $\psi(\mathbf{r},t=0)$, where $\psi_\sigma(\mathbf{r},t) = \langle \mathbf{r}\sigma|\psi(t)\rangle$ and $\sigma$ labels additional degrees of freedom associated for example with the electron spin or (in case of graphene) the pseudospin. Our objective is – in the same spirit of the pseudopotential theory used for calculating electronic band structures [14] – to obtain an effective Schrödinger equation satisfied by the smooth part of the wavefunction. To this end, it is convenient to introduce some form of spatial averaging, represented by a linear operator $\{\ \}_{av}$. The operator $\{\ \}_{av}$ is completely determined by the response function $F(\mathbf{k})$ such that $\{e^{i\mathbf{k}\cdot\mathbf{r}}\}_{av} = F(\mathbf{k})e^{i\mathbf{k}\cdot\mathbf{r}}$. In this work, we suppose that $\hat{H}$ describes a spatially-periodic system (e.g. a periodic superlattice) and assume that $\{\ \}_{av}$ corresponds to a low pass spatial filter, such that $F(\mathbf{k}) = 0$ for $\mathbf{k}$ outside the first Brillouin zone (B.Z.), and $F(\mathbf{k})=1$ otherwise. For example, if $\psi_\sigma(\mathbf{r},t)$ is a superposition of plane waves, $\psi_\sigma(\mathbf{r},t)=\sum_{\mathbf{k}} b_{\mathbf{k}} e^{i\mathbf{k}\cdot\mathbf{r}}$, then the macroscopic wavefunction resulting from spatial averaging is:

$$\{\psi_\sigma\}_{av}(\mathbf{r},t) = \sum_{\mathbf{k}\in B.Z.} b_{\mathbf{k}} e^{i\mathbf{k}\cdot\mathbf{r}}. \qquad (2)$$

Thus, only the spatial harmonics with $\mathbf{k}$ within the first Brillouin zone are retained after the spatial filtering. This type of spatial filtering is usually designated by ideal low-pass filtering.

By definition, a "macroscopic" state has the property $\{\psi(\mathbf{r})\}_{av} = \psi(\mathbf{r})$, i.e. a macroscopic state is unaffected by the spatial averaging. In particular, a "macroscopic state" cannot be more localized than the characteristic period of the material (or superlattice). As proven in Appendix A, a macroscopic state $|\psi\rangle$ is a superposition of states of the form $|\mathbf{k}\sigma\rangle$, with $\mathbf{k}$ in the first Brillouin zone. Similar definitions can be introduced in the context of electromagnetic metamaterials [12, 13], where the role of the macroscopic state is played by the macroscopic electromagnetic fields.

In Appendix A, it is shown that provided the initial state $|\psi(0)\rangle$ is macroscopic, then the smooth part of the wavefunction in the $|\mathbf{r}\sigma\rangle$ representation, $\Psi(\mathbf{r},t) \equiv \{\psi(\mathbf{r},t)\}_{av}$, satisfies exactly an homogenized Schrödinger equation of the form,

$$(\hat{H}_{ef}\Psi)(\mathbf{r},t) = i\hbar\frac{\partial}{\partial t}\Psi(\mathbf{r},t) \qquad (3)$$

where $\hat{H}_{ef}$ is the *effective Hamiltonian* of the system, defined in such a way that

$$\left(\hat{H}_{ef}\Psi\right)(\mathbf{r},t) = \left\{\left(\hat{H}\psi\right)(\mathbf{r},t)\right\}_{av} \quad \text{where} \quad \left(\hat{H}\psi\right)_{\sigma}(\mathbf{r},t) = \left\langle \mathbf{r}\sigma \,|\, \hat{H} \,|\, \psi \right\rangle.$$ The effective Hamiltonian can be written explicitly in terms of matrix elements $h_{\sigma,\sigma'}$, as follows

$$\left(\hat{H}_{ef}\Psi\right)_{\sigma} = \sum_{\sigma'} \int d^N\mathbf{r}' \int_0^t dt' \, h_{\sigma,\sigma'}(\mathbf{r}-\mathbf{r}', t-t') \Psi_{\sigma'}(\mathbf{r}',t'). \tag{4}$$

where $N$ is the dimension of the system (e.g. $N=2$ for a graphene sheet). Notice that the "microscopic" wavefunction is denoted by $\psi(\mathbf{r},t)$, whereas the macroscopic wavefunction (after spatial averaging) is denoted by $\Psi(\mathbf{r},t)$. The indicated properties of $h_{\sigma,\sigma'}(\mathbf{r},t)$ imply that its Fourier transform (unilateral in time and bilateral in space),

$$h_{\sigma,\sigma'}(\mathbf{k},\omega) = \int d^N\mathbf{r} \int_0^{+\infty} dt \, h_{\sigma,\sigma'}(\mathbf{r},t) e^{i\omega t} e^{-i\mathbf{k}\cdot\mathbf{r}}. \tag{5}$$

is such that

$$\left(\hat{H}_{ef}\Psi\right)_{\sigma}(\mathbf{k},\omega) = \sum_{\sigma'} h_{\sigma,\sigma'}(\mathbf{k},\omega) \Psi_{\sigma'}(\mathbf{k},\omega). \tag{6}$$

where $\Psi_{\sigma}(\mathbf{k},\omega) = \int d^N\mathbf{r} \int_0^{+\infty} dt \, \Psi_{\sigma}(\mathbf{r},t) e^{i\omega t} e^{-i\mathbf{k}\cdot\mathbf{r}}$ and $\left(\hat{H}_{ef}\Psi\right)_{\sigma}(\mathbf{k},\omega)$ is defined similarly (the convergence of the Fourier transform is guaranteed for $\text{Im}\{\omega\} > 0$). Hence, in the Fourier domain the relation between $\left(\hat{H}_{ef}\Psi\right)$ and $\Psi$ is a simple multiplication. It is interesting to mention that $h_{\sigma,\sigma'}(\mathbf{k},\omega)$ vanishes identically when $\mathbf{k}$ is outside the first Brillouin zone (see Appendix A), and this confirms that the effective Hamiltonian is indeed a "smooth operator" as compared to the original Hamiltonian.

Let us discuss how $h_{\sigma,\sigma'}(\mathbf{k},\omega)$ can be calculated in practice. As mentioned previously, the key property of the effective Hamiltonian is that if the initial state of system is "macroscopic", i.e. if $\{\psi(\mathbf{r},t=0)\}_{av} = \psi(\mathbf{r},t=0)$, then the result of averaging $(\hat{H}\psi)_\sigma = \langle \mathbf{r}\sigma|\hat{H}|\psi\rangle$ (with $|\psi\rangle$ the exact solution of the microscopic problem: $\hat{H}|\psi\rangle = i\hbar\frac{\partial}{\partial t}|\psi\rangle$) is exactly the same as that of applying the effective Hamiltonian to the averaged wavefunction $\Psi(\mathbf{r},t) = \{\psi(\mathbf{r},t)\}_{av}$. In particular, suppose that the initial state is such that $\psi_\sigma(\mathbf{r},t=0) = e^{i\mathbf{k}\cdot\mathbf{r}}\delta_{\sigma,\sigma'}$ and suppose that $\psi(\mathbf{r},t)$ is the corresponding exact solution of the Schrödinger equation (here $\mathbf{k}$ is fixed in the first Brillouin zone and $\sigma'$ is also fixed), and $\psi(\mathbf{r},\omega)$ is the corresponding (unilateral) Fourier transform in time. Taking into account that because of the assumed periodicity of the system in the spatial domain both $\psi(\mathbf{r},\omega)$ and $(\hat{H}\psi)_\sigma(\mathbf{r},\omega)$ must be Bloch modes associated with the wave vector $\mathbf{k}$, it is trivial to verify that $\{(\hat{H}\psi)_\sigma\}_{av}(\mathbf{r},\omega) = (\hat{H}\psi)_{\sigma,av} e^{i\mathbf{k}\cdot\mathbf{r}}$ and $\{\psi_\sigma\}_{av}(\mathbf{r},\omega) = \psi_{\sigma,av} e^{i\mathbf{k}\cdot\mathbf{r}}$, where

$$\psi_{\sigma,av}(\omega) = \frac{1}{V_{cell}} \int_\Omega d^N\mathbf{r}\, \psi_\sigma(\mathbf{r},\omega) e^{-i\mathbf{k}\cdot\mathbf{r}}, \tag{7a}$$

$$(\hat{H}\psi)_{\sigma,av}(\omega) = \frac{1}{V_{cell}} \int_\Omega d^N\mathbf{r}\, (\hat{H}\psi)_\sigma(\mathbf{r},\omega) e^{-i\mathbf{k}\cdot\mathbf{r}}. \tag{7b}$$

where $V_{cell}$ is the volume of the unit cell. Substituting these formulas into Eq. (4), we find that $\psi_{\sigma,av}(\omega)$ and $(\hat{H}\psi)_{\sigma,av}$ are linked by:

$$\left(\hat{H}\psi\right)_{\sigma,\mathrm{av}} = \sum_{\sigma'} h_{\sigma,\sigma'}\left(\mathbf{k},\omega\right)\psi_{\sigma',\mathrm{av}} \qquad (8)$$

This demonstrates that $h_{\sigma,\sigma'}(\mathbf{k},\omega)$ can be calculated numerically with the following algorithm: *(i)* solve the exact (time evolution) "microscopic" problem assuming an initial state such that $\psi_\sigma(\mathbf{r},t=0) \sim e^{i\mathbf{k}\cdot\mathbf{r}}\delta_{\sigma,\sigma'}$. *(ii)* determine $\psi(\mathbf{r},\omega)$ and $(\hat{H}\psi)(\mathbf{r},\omega)$, and afterwards $\psi_{\mathrm{av}}(\omega)$ and $(\hat{H}\psi)_{\mathrm{av}}$ defined consistently with Eq. (7). *(iii)* determine $h_{\sigma,\sigma'}(\mathbf{k},\omega)$ such that Eq. (8) is identically satisfied. In general, to obtain every component of $h_{\sigma,\sigma'}(\mathbf{k},\omega)$, one needs to solve several "microscopic" problems: as many as the degrees of freedom associated with $\sigma'$ in the initial time boundary condition $\psi_\sigma(\mathbf{r},t=0) \sim e^{i\mathbf{k}\cdot\mathbf{r}}\delta_{\sigma,\sigma'}$. It can be verified that the effective Hamiltonian ($h_{\sigma,\sigma'}(\mathbf{k},\omega)$) considered here in the context of matter waves is the analogue of the spatially dispersive effective dielectric function introduced in Refs. [12, 13] in the context of electromagnetic metamaterials. The explicit relation between the two formalisms is given in Appendix B. An important property of the effective Hamiltonian is that the corresponding energy eigenstates $E_n$, which satisfy for some non-trivial $\Psi$

$$\left(\hat{H}_{ef}\Psi\right)_{\omega=E_n/\hbar} = E_n\Psi, \qquad (9)$$

are exactly coincident with the energy eigenstates of the microscopic Hamiltonian. Therefore the electronic band structure of the microscopic Hamiltonian can be computed from the effective Hamiltonian. Strictly speaking, it should be mentioned that if an energy eigenvector of the microscopic Hamiltonian has a trivial projection into the subspace of macroscopic states, then the corresponding eigenvalue is not shared by $\hat{H}_{ef}$ and $\hat{H}$. This can only occur in degenerate singular cases, and for very specific forms of the microscopic Hamiltonian, and thus typically the electronic band structures of the microscopic and effective Hamiltonians are indeed the same. The enunciated properties are demonstrated in Appendix C.

## III. Graphene Superlattices

In the rest of the paper, we illustrate the application of the proposed homogenization method to the case of a graphene superlattice. Graphene is a one-atom-thick material whose low-energy excitations are massless, chiral, Dirac fermions [15]. Its unusual electronic properties make it a unique platform for the development of novel electronic devices with superior characteristics [16, 17, 18] and for flatland transformation optics [19]. Here, we will obtain an effective medium model for a graphene superlattice characterized by a 1D-electrostatic periodic potential, whose geometry is sketched in Fig. 1. Some recent works have shown that these graphene superlattices may be used to collimate an electron beam with virtually no spatial spreading or diffraction [20, 21]; quite differently here we concentrate on the effective medium description of electron waves in the superlattices, and highlight how by tailoring the microscopic potential one can control the "macroscopic" Hamiltonian of the quantum system.

The charge carriers in a graphene superlattice are described by the massless Dirac-type equation,

$$-i\hbar v_F (\boldsymbol{\sigma} \cdot \nabla)\psi + V(\mathbf{r})\psi = i\hbar \frac{\partial \psi}{\partial t}, \qquad (10)$$

where $v_F \sim 10^6 \, m/s$, $\boldsymbol{\sigma} = (\boldsymbol{\sigma}_x, \boldsymbol{\sigma}_y)$, $\boldsymbol{\sigma}_x, \boldsymbol{\sigma}_y$ are the Pauli matrices, and $V$ is the external periodic electrostatic potential. Here, without loss of generality, we restrict the analysis to the study of electrons whose transport properties are determined by the Dirac $K$ point [15]. The wavefunction $\psi$ has two components (it is a pseudo-spinor), each component, $\psi_m$ with $m=1,2$, being associated with a different trigonal sublattice of graphene. To obtain the effective Hamiltonian $h_{m,n}(\mathbf{k},\omega)$, $m,n=1,2$, it is sufficient to determine $\psi^{(l)}(\mathbf{r},\omega)$, i.e. the (unilateral) Fourier transform of the wavefunction $\psi^{(l)}(\mathbf{r},t)$ associated with an initial state of the form $\psi^{(l)}_m(\mathbf{r},t=0) = e^{i\mathbf{k}\cdot\mathbf{r}}\delta_{m,l}$ where $\mathbf{k}$ is measured with respect to the Dirac $K$ point and $l=1,2$. It should be noted that the electron spin plays no role in the absence of an external magnetic field, and hence $h_{m,n}(\mathbf{k},\omega)$ can be regarded a 2×2 matrix. As mentioned previously, the effective medium description is valid provided the initial electron state is not more localized than the characteristic period $a$ of the superlattice, i.e. provided $\mathbf{k}$ is inside the first Brillouin mini-zone of the superlattice.

From the properties of the (unilateral) Fourier transform, using $\frac{\partial \psi}{\partial t}(\mathbf{r},t) \leftrightarrow -i\omega\psi(\mathbf{r},\omega) - \psi_{t=0}(\mathbf{r})$, it is evident that $\psi^{(l)}(\mathbf{r},\omega)$ satisfies the time-independent equation:

$$\left[-i\hbar v_F(\boldsymbol{\sigma}\cdot\nabla) + V(\mathbf{r}) - \hbar\omega\right]\psi^{(l)} = -i\hbar\psi^{(l)}(\mathbf{r},t=0) \tag{11}$$

Notice that Eq. (11) is a non-homogeneous equation, where the initial time boundary condition $-i\hbar\psi^{(l)}(\mathbf{r},t=0)$ plays the role of a "source". On the other hand, it is straightforward to verify from Eq. (10) that $(\hat{H}\psi)_{av} = \hbar v_F(\boldsymbol{\sigma}\cdot\mathbf{k})\psi_{av} + (V\psi)_{av}$, where $(V\psi)_{av}$ is defined as $\psi_{av}$ in Eq. (7a), with $\psi$ replaced by $V\psi$. Hence, $h_{m,n}(\mathbf{k},\omega)$ is of the form,

$$h_{m,n}(\mathbf{k},\omega) = \left[\hbar v_F (\boldsymbol{\sigma} \cdot \mathbf{k})\right]_{m,n} + V_{ef,mn}(\mathbf{k},\omega) \qquad (12)$$

being $V_{ef,mn}(\mathbf{k},\omega)$ such that

$$(V\psi)_{m,av} = \sum_n V_{ef,mn}(\mathbf{k},\omega) \psi_{n,av} \ . \qquad (13)$$

Hence, for fixed $(\mathbf{k},\omega)$ the simplest way to determine the effective Hamiltonian is to solve Eq. (11) for $l=1,2$, and after this to compute the matrix $V_{ef}(\mathbf{k},\omega)$ such that

$$V_{ef}(\mathbf{k},\omega) = \left[(V\psi)_{av}^{(1)} ; (V\psi)_{av}^{(2)}\right] \cdot \left[\psi_{av}^{(1)} ; \psi_{av}^{(2)}\right]^{-1} . \qquad (14)$$

where the symbol ";" separates the columns of the 2×2 matrices. Evidently, in general, Eq. (11) needs to be solved numerically.

It is clear from the previous discussions, that in general $V_{ef}$ exhibits spatial dispersion (dependence on $\mathbf{k} \leftrightarrow -i\nabla$) and time dispersion (dependence on $\omega$, or equivalently on the energy $E = \hbar\omega$), and this introduces some complexity in the effective medium model. This is fully analogous to electromagnetic metamaterials, where in general the effective dielectric response depends on frequency and wave vector [12,13]. In order to further simplify the model, we consider the case where the spatial dispersion is weak so that it is possible to approximate $V_{ef}$ by its Taylor series of first order in $\mathbf{k}$,

$$V_{ef}(\mathbf{k},\omega) \approx V_{ef}(\omega) + \left.\frac{dV_{ef}}{dk_x}\right|_{(0,\omega)} k_x + \left.\frac{dV_{ef}}{dk_y}\right|_{(0,\omega)} k_y \quad \text{with} \quad V_{ef}(\omega) \equiv V_{ef}(0,\omega). \text{ Taking into}$$

account the symmetries of the microscopic Hamiltonian, it is simple to verify that $V_{ef}(\omega)$

is a scalar. Within this approximation, using Eqs. (4) and (12), it is seen that the effective Hamiltonian of the superlattice is such that:

$$\left(\hat{H}_{ef}\Psi\right)(\mathbf{r},\omega) = \left[-i\hbar v_F \boldsymbol{\sigma}_{ef}(\omega)\cdot\nabla + V_{ef}(\omega)\right]\cdot\Psi(\mathbf{r},\omega) \quad (15)$$

where $\Psi$ is a pseudo-spinor and $\boldsymbol{\sigma}_{ef}(\omega) \equiv (\boldsymbol{\sigma}_{x,ef}, \boldsymbol{\sigma}_{y,ef})$ with,

$$\boldsymbol{\sigma}_{ef,x} = \boldsymbol{\sigma}_x + \frac{1}{\hbar v_F}\frac{dV_{ef}}{dk_x}\bigg|_{(0,\omega)} \quad \text{and} \quad \boldsymbol{\sigma}_{ef,y} = \boldsymbol{\sigma}_y + \frac{1}{\hbar v_F}\frac{dV_{ef}}{dk_y}\bigg|_{(0,\omega)}. \quad (16)$$

In order to understand how the effective parameters vary with the energy, so that the model can be further simplified, in the next section we consider a numerical example.

## IV. Numerical Example

In Fig. 2, we depict the numerically calculated effective parameters as a function of $E = \hbar\omega$, for a superlattice characterized by a Krönig-Penney type electrostatic potential with $V_1 = -V_2$, $d_1 = d_2 = a/2$, and for the normalized potential amplitude $V_1 a / \hbar v_F = 6.0$. The effective potential tensor $V_{ef}(\mathbf{k},\omega)$ was calculated semi-analytically by solving Eq. (11) by matching plane wave modes at the interfaces between different regions, analogous to what is done when solving a scattering problem [22]. Very interestingly, the numerical results show that to an excellent approximation (Fig. 2b)

$$\boldsymbol{\sigma}_{ef} \approx v_{r,xx}(E)\boldsymbol{\sigma}_x \hat{\mathbf{x}} + v_{r,yy}(E)\boldsymbol{\sigma}_y \hat{\mathbf{y}}, \quad . \quad (17)$$

where $v_{r,ii}(E)$ are some scalars weakly dependent on $E$ and $\boldsymbol{\sigma}_i$ are the Pauli matrices. The pseudospin is characterized by $\boldsymbol{\sigma}_{ef}$ which can be written as $\boldsymbol{\sigma}_{ef} = \boldsymbol{\sigma} \cdot v_r$ with $v_r = v_{r,xx}\hat{\mathbf{x}}\hat{\mathbf{x}} + v_{r,yy}\hat{\mathbf{y}}\hat{\mathbf{y}}$. We shall refer to $v_r$ as the *anisotropy tensor* of the superlattice, and neglect its dependence on $E$ (due to its weak dependence on energy). On the other hand, the effective potential $V_{ef}$ (at $k=0$) for low-energy excitations varies linearly with $E$ so that,

$$V_{ef} \approx -\alpha E, \tag{18}$$

where $\alpha$ is a dimensionless positive constant. Notice that in pristine graphene $v_{r,ii}(E)=1$ and $\alpha=0$.

As discussed in Sect. II, the stationary states of the energy operator can be characterized using the developed effective medium approach, and the eigenvalues $E$ of the microscopic $\hat{H}$ Hamiltonian, are the same as the eigenvalues of the exact effective Hamiltonian $\hat{H}_{ef}$. In particular, within the validity of Eq. (15), the energy dispersion of the graphene superlattice at the Dirac $K$ point can be obtained by solving the eigenvalue problem

$$\left[-i\hbar v_F \boldsymbol{\sigma}_{ef} \cdot \nabla + V_{ef}(\omega)\right] \cdot \Psi = E\Psi. \tag{19}$$

For a spatial variation of the type $e^{i\mathbf{k}\cdot\mathbf{r}}$, this yields the dispersion $\left|E-V_{ef}\right| = \hbar v_F \sqrt{\left(v_{r,xx}k_x\right)^2 + \left(v_{r,yy}k_y\right)^2}$. In the case where $v_{r,ii}$ are independent of the energy and $V_{ef} \approx -\alpha E$, it is simple to check that the *group energy velocity* for propagation in the $x$-direction is: $v_{g,x} = \hbar^{-1}\partial_{k_x} E = v_F v_{r,xx}/(1+\alpha)$. On the other hand, because of the Klein tunneling effect [15], it is evident that $v_{g,x} = v_F$. This indicates that for this particular superlattice the effective parameters $v_{r,xx}$ and $\alpha$ are such that

$$v_{r,xx} = 1+\alpha. \tag{20}$$

We have indeed verified that the numerically calculated effective parameters satisfy exactly this relation. In particular, it follows that energy dispersion of the superlattice may be written simply as

$$|E| = \hbar v_F \sqrt{k_x^2 + \chi^2 k_y^2}, \qquad (21)$$

where $\chi = v_{r,yy}/v_{r,xx}$ is by definition the *anisotropy ratio*. The "exact" energy dispersion of a graphene superlattice with $V_1 a/\hbar v_F = 6.0$ is depicted in Fig. 3a, and is compared with our effective medium theory in Fig. 3b, revealing a very good agreement between both theories. As seen in Fig. 3a, consistent with the results of Ref. [21], the graphene superlattice is strongly anisotropic, and the usual Dirac cone of pristine graphene is stretched along the *y*-direction.

The pseudo-spinor associated with a state of energy $E$ and wave vector **k** is,

$$\Psi = \frac{1}{\sqrt{2}} \begin{pmatrix} 1 \\ se^{i\theta_q} \end{pmatrix} e^{i\mathbf{k}\cdot\mathbf{r}} \qquad (22)$$

where $\theta_\mathbf{q}$ is the angle between the vector $(v_{r,xx}k_x, v_{r,yy}k_y)$ and the *x*-axis and $s = \text{sgn}(E - V_{ef})$ [within the approximation $V_{ef} \approx -\alpha E$, we may also write that $s = \text{sgn}(E)$]. Thus, the pseudospin of the averaged wavefunction is determined by $\theta_\mathbf{q}$, and hence by the parameters $v_{r,ii}$. Since $v_{r,xx} = 1 + \alpha > 1$, the angle $\theta_\mathbf{q}$ may also be defined as the angle between $\mathbf{q} \equiv (k_x, \chi k_y)$ and the *x*-axis (Fig. 4b). Thus, in the case where the superlattice is characterized by strong anisotropy with $\chi \ll 1$ the pseudospin is such that either $\theta_\mathbf{q} \approx 0$ or $\theta_\mathbf{q} \approx \pi$. A similar result was reported in Ref. [21], based on direct band structure calculations. It should be noted that, unlike in pristine graphene, $\theta_\mathbf{q} \neq \theta_\mathbf{k}$, where $\theta_\mathbf{k}$ is the angle between $\mathbf{k}$ and the *x*-axis (Fig. 4b). Evidently, the angle $\theta_\mathbf{q}$ determines the relative phase of the wavefunction in the two sublattices of graphene.

In Fig. 2c-d we plot $v_{r,ii}$ and $\chi$ (evaluated for $E \approx 0$), respectively, as a function of the normalized potential $V_1$. As seen, the anisotropy ratio can be quite large if $V_1 a / \hbar v_F \sim 2\pi$, which is when the parameter $v_{r,yy}$ crosses zero. This extreme regime of operation may permit a supercollimation of an electron beam [21].

## V. Perfect Tunneling

One of the most striking features of Fig. 2d is the fact that $\chi$ can be negative for an applied potential with normalized amplitude $V_1 a / \hbar v_F > 6.3$. What are the physical consequences of a reversed sign for the anisotropy ratio? To answer this question we consider two graphene based nanomaterials described by the effective parameters $\chi_1$ and $\chi_2$ such that $\chi_1 = -\chi_2 \equiv \chi$ and $\chi_1 > 0$, where $\chi_i = \left(v_{r,yy} / v_{r,xx}\right)_i$. At a microscopic level these nanomaterials may be regarded as superlattices characterized by a suitable microscopic potential, consistent with the previous discussions. Moreover, we assume that there is a static potential offset $\delta V$ between the nanomaterials, so that the energy dispersion of the stationary modes in the first material is $|E| = \hbar v_F \sqrt{k_x^2 + \chi^2 k_y^2}$, whereas the dispersion in the second material is $|E - \delta V| = \hbar v_F \sqrt{k_x^2 + \chi^2 k_y^2}$. We say that the pseudospin of a material is *positive* when $\chi_i > 0$ and that the pseudospin is *negative* if $\chi_i < 0$. Consider now the geometry depicted in Fig. 4d, which shows a rectangular slab of the second material embedded in the first material. Using the developed effective medium theory, it is straightforward to compute the transmissivity for a stationary state (plane wave) that describes an electron incident from the left region, $x < 0$, on the slab of the material with negative anisotropy. This is done by expanding the wave function in the different regions into plane wave modes, and ensuring the continuity of the pseudo-spinors at the interfaces (see Appendix D). In Fig. 4c the transmissivity is shown as a function of the transverse wave number ($k_y$) of the incoming particle, for different values of the energy. Quite interestingly, it is seen that there is a perfect tunneling – independent of the direction of incidence of the incoming particle – when the energy is $E = \delta V / 2$ (green curve). This is distinctively different from Klein tunneling [15], which only occurs for $k_y = 0$, whereas in our case the incident electron tunnels through the second material, independent of the angle of incidence! The tunneling phenomenon at $E = \delta V / 2$ can be easily understood on the basis of the effective medium theory. Indeed, taking into account that the energy level $E = \delta V / 2$ is in the conduction band of the first material and in the valence band of the second material (Fig. 4a), and that for such a value of $E$ the energy dispersions of both materials are coincident, it follows that if the pseudo-spinor of the incident wave is $\Psi_1 \sim \begin{pmatrix} 1 \\ e^{i\theta_{q,1}} \end{pmatrix} e^{ik_y y} e^{ik_x x}$ then the pseudo-spinor of the wave that propagates in the second material along the positive $x$-direction is $\Psi_2 \sim \begin{pmatrix} 1 \\ -e^{i\theta_{q,2}} \end{pmatrix} e^{ik_y y} e^{-ik_x x}$.

But since $\chi_1 = -\chi_2 \equiv \chi$ it is evident that $\theta_{q,1} = \theta_{q,2} + \pi$, and hence $\begin{pmatrix} 1 \\ e^{i\theta_{q,1}} \end{pmatrix}$ and $\begin{pmatrix} 1 \\ -e^{i\theta_{q,2}} \end{pmatrix}$ are equal (see Fig. 4b). This implies perfect matching at the interfaces, and hence perfect transmission independent of the angle of incidence. It is interesting to point out that there is an evident parallelism between our graphene-based nanomaterial with negative pseudospin and double negative (DNG) electromagnetic metamaterials, as proposed in [4]. Indeed, similar to electromagnetic metamaterials, our graphene-based nanomaterial may provide perfect focusing and negative refraction of electrons with the energy $E = \delta V/2$, when the source of electrons is at a distance $W/2$ from the slab of the second material (Fig. 4d). A graphene analogue of Veselago-Pendry's lens has been proposed in an earlier publication [17], however very different from the configuration considered here, a p-n graphene junction can only mimic Veselago-Pendry's lens under a semi-classical approximation because the electron transmissivity there is very different from unity for wide incident angles. Moreover, our solution here behaves as a perfect lens even for incoming electron waves characterized by a complex wave vector $\mathbf{k}_1 = (k_{x1}, k_y)$, i.e. for states such that $k_y > k_{y,\max}$ where $k_{y,\max} = E/(\chi_1 \hbar v_F)$ is the transverse wave vector for grazing incidence. For such complex states, which strictly speaking are non-normalizable, $k_{x1}$ is pure imaginary, and the transmissivity of the structure is exactly $|T|^2 = |e^{-i2k_{x1}W}| > 1$ (growing exponential: see Fig. 4c), which similarly to Pendry's lens compensates for the exponential decay in the outside regions [4]. Even though the mentioned complex states are non-normalizable this result is full of physical significance. For example, consider the stationary states in the scenario where an arbitrary localized external perturbation (e.g. a potential well or barrier somewhere in the region $x < 0$) is introduced in the first material. Since two materials with opposite pseudospin completely annihilate one another when $E = \delta V/2$, this means that a pair of such materials may be "inserted" into the considered structure in the region $x > 0$ without changing in any manner the stationary states associated with $E = \delta V / 2$ in the region $x < 0$.

It is natural to wonder if the material with positive pseudospin in the scenario of Fig. 4 may be chosen as pristine graphene. This requires that $\chi_1 = 1$ and thus $\chi_2 = -1$. Unfortunately, it can be checked in Fig. 2c that the minimum value for $1/|\chi|$, with $\chi < 0$, for a superlattice characterized by a Kronig-Penney type potential is about $4.6$. However, this does not preclude that for a different profile of the microscopic potential, $1/|\chi|$ cannot be made smaller. For example, by solving the effective medium problem numerically using a finite-difference method, we calculated $1/\chi$ as a function of $V_1$ for a microscopic potential of the form $V(x) = V_1 \sin\left(\frac{2\pi}{a} x\right)$ (the unit cell is discretized as $N \times N$ array of nodes and the derivatives are replaced by finite differences; then the problem is reduced to the solution of a linear system). The obtained result is represented in Fig. 2d with a dashed line. As seen, for this alternative potential the minimum value of $1/|\chi|$ (with $\chi < 0$) is reduced to $2.5$. This is still far from $\chi_2 = -1$, but indicates that by tailoring the shape of the microscopic potential it may at least be possible to better approximate the value $\chi_2 = -1$.

## VI. Conclusion

In conclusion, we have described a completely general self-consistent approach (many body effects are however neglected) to characterize electron waves in periodic systems from an effective medium perspective, which extends our previous work on electromagnetic metamaterials to matter waves [12, 13]. The proposed theory may be instrumental to establish novel analogies between electromagnetics and electronics, as further pursued in Ref. [24] in case of semiconductor superlattices. Here, we applied the formalism to the case of a graphene superlattice characterized by a 1D-periodic potential, showing that the low energy excitations can be described in terms of an effective scalar potential and an anisotropy tensor $v_r = v_{r,xx}\hat{\mathbf{x}}\hat{\mathbf{x}} + v_{r,yy}\hat{\mathbf{y}}\hat{\mathbf{y}}$. In particular, based on our effective medium model we predict a regime of perfect tunneling between nanomaterials with symmetric values of the anisotropy ratio $v_{r,yy}/v_{r,xx}$ and for a specific value of the electron energy, completely analogous to Pendry's perfect lens in the context of electromagnetic metamaterials.

This work is supported in part by the U.S. Air Force of Scientific Research (AFOSR) grant numbers FA9550-08-1-0220 and FA9550-10-1-0408, and by Fundação para a Ciência e a Tecnologia grant number PTDC/EEATEL/100245/2008.

## Appendix A: The effective Schrödinger equation

The spatial averaging operator defined in the main text in the coordinate representation, $\{\ \}_{av}$, can more generally be described by the projection operator:

$$\hat{O}_{av} = \frac{1}{(2\pi)^N} \sum_\sigma \int_{B.Z.} |\mathbf{k}\sigma\rangle\langle\mathbf{k}\sigma| d^N\mathbf{k} \tag{A1}$$

where B.Z. represents the first Brillouin zone, and the normalization $\langle\mathbf{k}'|\mathbf{k}\rangle = (2\pi)^N \delta(\mathbf{k}'-\mathbf{k})$ is implicit (here, $|\mathbf{k}\rangle$ represents a state such that $\langle\mathbf{r}|\mathbf{k}\rangle = e^{i\mathbf{k}\cdot\mathbf{r}}$). The label $\sigma$ represents additional degrees of freedom of the wavefunction, such as the

electron spin and/or (in case of graphene) the pseudo-spin. Indeed it is simple to check that $\{\psi_\sigma(\mathbf{r},t)\}_{av} = \langle\mathbf{r}\sigma|\hat{O}_{av}|\psi(t)\rangle$.

It is also useful to introduce the (unilateral) Fourier transform of $|\psi(t)\rangle$ given by

$$|\psi(\omega)\rangle = \int_0^{+\infty} dt |\psi(t)\rangle e^{+i\omega t} dt. \tag{A2}$$

The Fourier transform is defined for $\text{Im}(\omega) > 0$. The Fourier transform of other state vectors or time dependent operators is defined similarly.

We denote the averaged state vector of the system as $|\psi_{av}(t)\rangle \equiv \hat{O}_{av}|\psi(t)\rangle$, and we define $|(\hat{H}\psi)_{av}(t)\rangle \equiv \hat{O}_{av}\hat{H}|\psi(t)\rangle$. Here, the objective is to find an effective Hamiltonian that links $|(\hat{H}\psi)_{av}(t)\rangle$ and $|\psi_{av}(t)\rangle$, in such a way that $\hat{H}_{ef}(\omega)|\psi_{av}(\omega)\rangle = |(\hat{H}\psi)_{av}(\omega)\rangle$ in the frequency domain, when the initial state of the system is macroscopic, i.e. when $|\psi_{t=0}\rangle = |(\psi_{t=0})_{av}\rangle$.

Denoting the propagator of the system by $\hat{U}(t)$, we can write:

$$|\psi_{av}(t)\rangle = \hat{O}_{av}\hat{U}(t)|\psi_{t=0}\rangle \quad ; \quad |(\hat{H}\psi)_{av}(t)\rangle = \hat{O}_{av}\hat{G}(t)|\psi_{t=0}\rangle \tag{A3}$$

where $\hat{G} = \hat{H}\hat{U}$. In the Fourier domain these relations are equivalent to:

$$|\psi_{av}(\omega)\rangle = \hat{O}_{av}\hat{U}(\omega)|\psi_{t=0}\rangle \quad ; \quad |(\hat{H}\psi)_{av}(\omega)\rangle = \hat{O}_{av}\hat{G}(\omega)|\psi_{t=0}\rangle \tag{A4}$$

If the initial state is macroscopic, i.e. $|\psi_{t=0}\rangle = |(\psi_{t=0})_{av}\rangle$, it is evident that:

$$|\psi_{av}(\omega)\rangle = \hat{O}_{av}\hat{U}(\omega)\hat{O}_{av}|\psi_{t=0}\rangle \tag{A5}$$

The operator $\hat{O}_{av}\hat{U}(\omega)\hat{O}_{av}$ maps the subspace of "macroscopic" wavefunctions onto itself. Let $\left[\hat{O}_{av}\hat{U}(\omega)\hat{O}_{av}\right]^{-1}$ represent its inverse in this subspace. Then, we have:

$$\left|\left(\hat{H}\psi\right)_{av}(\omega)\right\rangle = \hat{H}_{ef}(\omega)\left|\psi_{av}(\omega)\right\rangle \tag{A6}$$

where $\hat{H}_{ef}(\omega) = \hat{O}_{av}\hat{G}(\omega)\left[\hat{O}_{av}\hat{U}(\omega)\hat{O}_{av}\right]^{-1}\hat{O}_{av}$. Thus, $\hat{H}_{ef}(\omega)$ is the desired effective Hamiltonian of the quantum system. It should be underlined that the above formula holds exactly, provided the initial state of the system (at *t=0*) is macroscopic.

If $\hat{H}_{ef}(t)$ is the inverse Fourier transform of $\hat{H}_{ef}(\omega)$ (with $\hat{H}_{ef}(t) = 0$ for $t < 0$), i.e. if

$$\hat{H}_{ef}(\omega) = \int_0^{+\infty} dt\, \hat{H}_{ef}(t) e^{+i\omega t} dt, \tag{A7}$$

it is possible to write in the time domain

$$\left|\left(\hat{H}\psi\right)_{av}(t)\right\rangle = \int_0^t dt'\, \hat{H}_{ef}(t-t')\left|\psi_{av}(t')\right\rangle. \tag{A8}$$

It is worth mentioning that if the integration region in the right-hand side of Eq. (A1) were taken as the entire momentum space, then $\hat{O}_{av} = \hat{I}$. In that case, if $\hat{H}$ is independent of time, we would obtain $\hat{H}_{ef}(t) = \delta(t)\hat{H}$. The role of $\hat{O}_{av}$ in the definition of $\hat{H}_{ef}$ is thus to "smooth" the exact "microscopic" Hamiltonian.

Since the "microscopic" wavefunction satisfies the Schrödinger equation, $\hat{H}|\psi\rangle = i\hbar\frac{\partial}{\partial t}|\psi\rangle$, it is evident that provided the initial state (at *t=0*) is macroscopic, then average state vector $|\psi_{av}(t)\rangle$ satisfies exactly:

$$\int_0^t dt' \hat{H}_{ef}(t-t') |\psi_{av}(t')\rangle = i\hbar \frac{\partial}{\partial t} |\psi_{av}(t)\rangle. \tag{A9}$$

Notice that in this effective medium description the action of the Hamiltonian at time $t$ depends explicitly on the past history of the state vector, i.e. on the values of $|\psi_{av}(t')\rangle$ for $t' < t$ (note, however, that the past history can be traced back to the value of the wave function at time $t = 0$, using Eq. (A9)). This intrinsic "time dispersion" in the response is the price that is paid for the effective medium description of the system.

Now that we have a formal definition for the effective Hamiltonian $\hat{H}_{ef}(\omega)$, we want to obtain its representation in the momentum space. To this end, we calculate,

$$H_{ef}(\mathbf{k}\sigma, \mathbf{k}'\sigma'; \omega) = \langle \mathbf{k}\sigma | \hat{H}_{ef}(\omega) | \mathbf{k}'\sigma' \rangle \tag{A10}$$

being $\hat{H}_{ef}(\omega) = \hat{O}_{av} \hat{G}(\omega) \left[ \hat{O}_{av} \hat{U}(\omega) \hat{O}_{av} \right]^{-1} \hat{O}_{av}$. Now the key point is that because the quantum system is assumed invariant to translations along the basis vectors of the direct lattice, it follows that $\langle \mathbf{k}\sigma | \hat{G}(\omega) | \mathbf{q}s \rangle$ and $\langle \mathbf{k}\sigma | \hat{U}(\omega) | \mathbf{q}s \rangle$ vanish except if $\mathbf{k} - \mathbf{q}$ is a primitive vector of the reciprocal lattice. In particular, it is evident that $\hat{O}_{av} \hat{U}(\omega) \hat{O}_{av}$ maps the state $|\mathbf{k}\sigma\rangle$ into a state of the form $\sum_s c_s |\mathbf{k}s\rangle$, and hence the inverse function has the same property. This implies that $\hat{H}_{ef}(\omega)$ also maps the state $|\mathbf{k}\sigma\rangle$ into a state of the form $\sum_s c_s |\mathbf{k}s\rangle$, and thus it follows that:

$$H_{ef}(\mathbf{k}\sigma, \mathbf{k}'\sigma'; \omega) = h_{\sigma,\sigma'}(\mathbf{k}, \omega) (2\pi)^N \delta(\mathbf{k} - \mathbf{k}') \tag{A11}$$

where $h_{\sigma,\sigma'}(\mathbf{k}, \omega) = \int \frac{d^N \mathbf{k}'}{(2\pi)^N} \langle \mathbf{k}\sigma | \hat{H}_{ef}(\omega) | \mathbf{k}'\sigma' \rangle$. In particular, this result implies that:

$$\hat{H}_{ef}(\omega) = \sum_{\sigma,\sigma'} \int \frac{d^N \mathbf{k}'}{(2\pi)^N} \frac{d^N \mathbf{k}}{(2\pi)^N} |\mathbf{k}\sigma\rangle\langle \mathbf{k}\sigma | \hat{H}_{ef}(\omega) | \mathbf{k}'\sigma'\rangle\langle \mathbf{k}'\sigma'|$$
$$= \sum_{\sigma,\sigma'} \int \frac{d^N \mathbf{k}}{(2\pi)^N} h_{\sigma,\sigma'}(\mathbf{k},\omega) |\mathbf{k}\sigma\rangle\langle \mathbf{k}\sigma'| \qquad (A12)$$

It is worth noting that $h_{\sigma,\sigma'}(\mathbf{k},\omega)$ vanishes for $\mathbf{k}$ outside the Brillouin zone. The above formula gives the desired representation of $\hat{H}_{ef}(\omega)$ in the momentum basis. Substituting this result into Eq. (A6), it is found that,

$$\langle \mathbf{k}\sigma | (\hat{H}\psi)_{av}(\omega) \rangle = \sum_{\sigma'} h_{\sigma,\sigma'}(\mathbf{k},\omega) \langle \mathbf{k}\sigma' | \psi_{av}(\omega) \rangle \qquad (A13)$$

i.e. in the momentum coordinates the application of $\hat{H}_{ef}(\omega)$ reduces to a simple matrix multiplication. Equivalently, we can also write:

$$\langle (\hat{H}\psi)_{\sigma} \rangle (\mathbf{r},t) = \sum_{\sigma'} \int d^N \mathbf{r}' \int_0^t dt' \, h_{\sigma,\sigma'}(\mathbf{r}-\mathbf{r}', t-t') \Psi_{\sigma'}(\mathbf{r}',t') \qquad (A14)$$

where $h_{\sigma,\sigma'}(\mathbf{r},t)$ is the inverse transform of $h_{\sigma,\sigma'}(\mathbf{k},\omega)$, i.e. $h_{\sigma,\sigma'}(\mathbf{k},\omega) = \int d^N \mathbf{r} \int_0^{+\infty} dt \, h_{\sigma,\sigma'}(\mathbf{r},t) e^{i\omega t} e^{-i\mathbf{k}\cdot\mathbf{r}}$, and by definition $\Psi_{\sigma}(\mathbf{r},t) = \langle \mathbf{r}\sigma | \psi_{av}(t) \rangle$ and $\langle (\hat{H}\psi)_{\sigma} \rangle (\mathbf{r},t) = \langle \mathbf{r}\sigma | (\hat{H}\psi)_{av}(t) \rangle$.

From Eq. (A9) it is thus evident that if the initial state (at $t=0$) is macroscopic, then the "macroscopic" wavefunction $\Psi_{\sigma}(\mathbf{r},t)$ satisfies exactly:

$$\sum_{\sigma'} \int d^N \mathbf{r}' \int_0^t dt' \, h_{\sigma,\sigma'}(\mathbf{r}-\mathbf{r}', t-t') \Psi_{\sigma'}(\mathbf{r}',t') = i\hbar \frac{\partial}{\partial t} \Psi_{\sigma}(\mathbf{r},t) \qquad (A15)$$

In other words, provided the initial state is macroscopic the time evolution of the macroscopic wavefunction ($|\psi_{av}(t)\rangle = \hat{O}_{av}|\psi(t)\rangle$) is fully determined by the effective Hamiltonian $\hat{H}_{ef}(\omega)$, through a modified Schrödinger-type equation, as in Eq. (A15).

## Appendix B: Effective medium description of electromagnetic metamaterials

In this Appendix, we discuss the explicit connection between the formalism developed in the main text and that of our previous work on electromagnetic metamaterials, highlighting the equivalence between the two [12,13]. The starting point is to note that the Maxwell's equations in a continuous medium can be written as,

$$\begin{pmatrix} 0 & i\nabla\times \\ -i\nabla\times & 0 \end{pmatrix} \mathbf{f} = i\frac{\partial \mathbf{g}}{\partial t} \tag{B1}$$

with $\mathbf{f} = \begin{pmatrix} \mathbf{e} \\ \mathbf{h} \end{pmatrix}$ and $\mathbf{g} = \begin{pmatrix} \mathbf{d} \\ \mathbf{b} \end{pmatrix}$. For standard isotropic non-dispersive magneto-dielectrics, the electric and magnetic fields, $\mathbf{e}$ and $\mathbf{h}$, are linked to the electric displacement and magnetic induction fields, $\mathbf{d}$ and $\mathbf{b}$, by the standard constitutive relations $\mathbf{g} = \mathbf{M}\cdot\mathbf{f}$ with $\mathbf{M} = \begin{pmatrix} \varepsilon & 0 \\ 0 & \mu \end{pmatrix}$. For simplicity, in the following discussion we neglect material dispersion.

In a periodic metamaterial the permittivity and permeability are periodic functions of space: $\varepsilon = \varepsilon(\mathbf{r})$ and $\mu = \mu(\mathbf{r})$. Hence, the dynamics of the "microscopic" electromagnetic fields (i.e. before any form of averaging on the scale of the unit cell of the metamaterial) can be described by a Schrödinger-type equation of the form $\hat{H}\psi = i\hbar\frac{\partial}{\partial t}\psi$ where $\psi \leftrightarrow \mathbf{g}$ is a six-component vector, and the operator $\hat{H}$ is given by:

$$\hat{H} = \hbar \begin{pmatrix} 0 & i\nabla \times \\ -i\nabla \times & 0 \end{pmatrix} \cdot \mathbf{M}^{-1}. \tag{B2}$$

Obviously, the above Hamiltonian does not represent the energy of the system, but it is rather an operator that describes the dynamics of the classical electromagnetic field. For simplicity, from now on in this Appendix we set $\hbar = 1$. It is interesting to notice that $\hat{H}$ is Hermitian with respect to the inner product,

$$\langle \mathbf{g}_2 | \mathbf{g}_1 \rangle = \frac{1}{2} \int d^3\mathbf{r}\, \mathbf{g}_2^* \cdot \mathbf{M}^{-1}(\mathbf{r}) \cdot \mathbf{g}_1, \tag{B3}$$

being implicit that the six-vector fields satisfy suitable (e.g. periodic) boundary conditions at the boundary of the integration region. We also note that $\langle \mathbf{g} | \mathbf{g} \rangle$ is the stored electromagnetic energy.

For a given initial state of the electromagnetic field $\mathbf{g}_{t=0}$, the unilateral Fourier transform in time of $\mathbf{g}$, defined by $\tilde{\mathbf{g}} = \int_0^{+\infty} \mathbf{g} e^{i\omega t} dt$, satisfies:

$$\hat{H}\tilde{\mathbf{g}} = \omega \tilde{\mathbf{g}} - i\mathbf{g}_{t=0} \tag{B4}$$

The above result implies that for initial macroscopic states of the form $\mathbf{g}_{t=0} = \mathbf{g}_0 e^{i\mathbf{k}\cdot\mathbf{r}}$, with $\mathbf{g}_0$ being an arbitrary constant vector, the macroscopic fields satisfy $(\hat{H}\tilde{\mathbf{g}})_{av} = \omega \mathbf{G} - i\mathbf{g}_0$, where $\mathbf{G} = \tilde{\mathbf{g}}_{av} = \frac{1}{V_{cell}} \int_\Omega \tilde{\mathbf{g}} e^{-i\mathbf{k}\cdot\mathbf{r}} d^3\mathbf{r}$ represents the macroscopic electric displacement and macroscopic induction fields $\mathbf{G} = \begin{pmatrix} \mathbf{D} \\ \mathbf{B} \end{pmatrix}$, $(\hat{H}\tilde{\mathbf{g}})_{av} = \frac{1}{V_{cell}} \int_\Omega (\hat{H}\tilde{\mathbf{g}}) e^{-i\mathbf{k}\cdot\mathbf{r}} d^3\mathbf{r}$, $\Omega$ is the unit cell and $V_{cell}$ is the corresponding volume. From the theory of the main text, the effective medium operator $\hat{H}_{ef}(\omega, \mathbf{k})$ is determined by imposing that for arbitrary $\mathbf{g}_0$ one has:

$$\hat{H}_{ef}\mathbf{G} = \left(\hat{H}\tilde{\mathbf{g}}\right)_{av}. \tag{B5}$$

It is easy to check from the definition of $\hat{H}$ that

$$\left(\hat{H}\tilde{\mathbf{g}}\right)_{av} = \begin{pmatrix} 0 & -\mathbf{k}\times \\ \mathbf{k}\times & 0 \end{pmatrix} \cdot \mathbf{F} \tag{B6}$$

where $\mathbf{F} = \begin{pmatrix} \mathbf{E} \\ \mathbf{H} \end{pmatrix}$ represents the macroscopic electric and magnetic fields, defined so that

$\mathbf{F} = \dfrac{1}{V_{cell}} \int_{\Omega} \left(\mathbf{M}^{-1} \cdot \tilde{\mathbf{g}}\right) e^{-i\mathbf{k}\cdot\mathbf{r}} d^3\mathbf{r}$. Let the matrix $\mathbf{M}_{ef}(\omega, \mathbf{k})$ be such that for arbitrary $\mathbf{g}_0$ (i.e. for an arbitrary initial macroscopic state), one has:

$$\mathbf{G} = \mathbf{M}_{ef}(\omega, \mathbf{k}) \cdot \mathbf{F}. \tag{B7}$$

It is easy to check that such a matrix exists and is uniquely defined, and the above equation is actually the basis of the homogenization approach of Refs [12, 13]. It is important to mention, as already discussed in Ref. [13], that defining $\mathbf{M}_{ef}(\omega, \mathbf{k})$ such that Eq. (B7) holds for an arbitrary initial macroscopic state, as considered here, is fully equivalent to define $\mathbf{M}_{ef}(\omega, \mathbf{k})$ in such a way that Eq. (B7) holds for an arbitrary external macroscopic harmonic current excitation, which is the original framework of Ref. [12].

From Eq. (B5) and Eq. (B7), the effective medium operator can be expressed as:

$$\hat{H}_{ef} = \begin{pmatrix} 0 & -\mathbf{k}\times \\ \mathbf{k}\times & 0 \end{pmatrix} \cdot \mathbf{M}_{ef}^{-1}(\omega, \mathbf{k}). \tag{B8}$$

The matrix $\mathbf{M}_{ef}(\omega, \mathbf{k})$ gives the effective medium parameters of the metamaterial, and is consistent with the definition of Ref. [10]. In the particular case of a metamaterial formed by non-magnetic particles ($\mu = \mu_0$) it can be shown that $\mathbf{M}_{ef}(\omega, \mathbf{k})$ assumes the simple form $\mathbf{M}_{ef}(\omega, \mathbf{k}) = \begin{pmatrix} \bar{\bar{\varepsilon}}_{ef} & 0 \\ 0 & \mu_0 \end{pmatrix}$ where $\bar{\bar{\varepsilon}}_{ef} = \bar{\bar{\varepsilon}}_{ef}(\omega, \mathbf{k})$ is the nonlocal effective dielectric function, defined exactly as in our previous works [12, 13]. This confirms that the theory of this work generalizes the previous studies [12, 13].

## Appendix C: The stationary states

The energy eigenstates of a microscopic Hamiltonian $\hat{H}$ (with $\hat{H}$ independent of time) are exactly the same as those of the corresponding homogenized system described by $\hat{H}_{ef}$, except for some possible degenerate cases that are discussed below. This result can be proven by noting that the time evolution of the wavefunction for a given initial state $|\psi_{t=0}\rangle$ is evidently $|\psi(t)\rangle = \sum_n c_n |n\rangle e^{-i\frac{E_n}{\hbar}t}$ with $c_n = \langle n | \psi_{t=0} \rangle$, with $|n\rangle$, $n=1,2,\ldots$, being the energy eigenstates of the microscopic system and $E_n$ the corresponding eigen-energies. In particular it is evident that $|\psi(\omega)\rangle$, defined as in Eq. (A2) for $\mathrm{Im}\,\omega > 0$, is

$$|\psi(\omega)\rangle = \sum_n c_n |n\rangle \frac{1}{i(\omega_n - \omega)}, \text{ with } \omega_n = E_n / \hbar.$$ On the other hand, from the definition of $\hat{H}_{ef}(\omega)$ [see Eq. (A6)], we know that if $|\psi_{t=0}\rangle$ is a macroscopic state, then

$$\hat{O}_{av} \hat{H} |\psi(\omega)\rangle = \hat{H}_{ef}(\omega) \hat{O}_{av} |\psi(\omega)\rangle \tag{C1}$$

Hence, this implies that:

$$\sum_n c_n E_n |n_{av}\rangle \frac{1}{i(\omega_n - \omega)} = \sum_n c_n \hat{H}_{ef}(\omega) |n_{av}\rangle \frac{1}{i(\omega_n - \omega)} \tag{C2}$$

where we put $|n_{av}\rangle = \hat{O}_{av} |n\rangle$. In order that to have an identity, it is necessary that the residues of both sides of the equation calculated at a generic pole, $\omega = \omega_n$, are equal. Hence, it follows that

$$\hat{H}_{ef}(\omega_n) |n_{av}\rangle = E_n |n_{av}\rangle, \quad \text{with } \omega_n = E_n / \hbar. \tag{C3}$$

This confirms that the energy eigenstates of a microscopic Hamiltonian $\hat{H}$, should be precisely the same as those of the homogenized system. The eigenstates of the homogenized system are evidently $|n_{av}\rangle = \hat{O}_{av} |n\rangle$. Strictly speaking, some degenerate cases for which this property does not hold may occur. This may happen only if $|n_{av}\rangle = 0$, i.e. in case of microscopic states that have a trivial projection into the subspace of macroscopic states. In such a case Eq. (C3) is equivalent to $0 = 0$ and hence $\hat{H}_{ef}(\omega_n)$ is not required to have a non-trivial null space. In such circumstances the spectra of the microscopic and macroscopic Hamiltonians may not be exactly coincident at a few isolated points. The states for which $|n_{av}\rangle = 0$, if there are any, cannot be excited with an initial state $|\psi_{t=0}\rangle$ that is macroscopic.

It is important to mention that the eigenstates of the homogenized system are not, in general, mutually orthogonal with respect to the scalar product of the original Hilbert space. In fact, we have $\langle m_{av} | n_{av} \rangle = \langle m | \hat{O}_{av} | n \rangle$, and in general $\langle m | \hat{O}_{av} | n \rangle \neq \langle m | n \rangle = \delta_{m,n}$. Moreover, if $|\psi_{av}(t)\rangle$ is solution of the macroscopic

Schrödinger equation [Eq. (A15)] (for a given initial time macroscopic state $|\psi_{t=0}\rangle$), then in general $\langle\psi_{av}(t)|\psi_{av}(t)\rangle = \langle\psi(t)|\hat{O}_{av}|\psi(t)\rangle$ may be different from $\langle\psi(t)|\psi(t)\rangle = 1$. In other words, the standard normalization of the wavefunction does not apply to the averaged state vector.

## Appendix D: Electron tunneling through a nanomaterial with a negative pseudospin

Here, we describe the model used to compute the transmission coefficient when an electron with energy $E$ propagating on a graphene based nanomaterial with the energy dispersion $E = +\hbar v_F \sqrt{k_x^2 + \chi_1^2 k_y^2}$ (conduction band), impinges on another graphene nanomaterial (with thickness $W$ along $x$) with the energy dispersion $E = \delta V - \hbar v_F \sqrt{k_x^2 + \chi_2^2 k_y^2}$ (valence band). The anisotropy ratio is $\chi_i = (v_{r,yy}/v_{r,xx})_i$.

We assume that the Dirac points in the nanomaterials ($K$ and $K'$ in case of pristine graphene; in a graphene superlattice extra Dirac points may emerge [22, 23]) may be regarded as independent and associated with different physical channels. We are only interested in the scattering of electrons with wave vector close to the Dirac $K$ point. Thus, the electron pseudo-spinor in the nanomaterials may be written as (the wave vector is measured with respect to the $K$ point):

$$\Psi = \begin{pmatrix} 1 \\ e^{i\theta_{q,1}} \end{pmatrix} e^{ik_y y} e^{ik_{x1} x} + R \begin{pmatrix} 1 \\ -e^{-i\theta_{q,1}} \end{pmatrix} e^{ik_y y} e^{-ik_{x1} x}, \quad x < 0 \tag{D1}$$

$$\Psi = A \begin{pmatrix} 1 \\ -e^{i\theta_{q,2}} \end{pmatrix} e^{ik_y y} e^{ik_{x2} x} + B \begin{pmatrix} 1 \\ e^{-i\theta_{q,2}} \end{pmatrix} e^{ik_y y} e^{-ik_{x2} x}, \quad 0 < x < W \tag{D2}$$

$$\Psi = T \begin{pmatrix} 1 \\ e^{i\theta_{q,1}} \end{pmatrix} e^{ik_y y} e^{ik_{x1} x}, \qquad x > W \tag{D3}$$

where $R$ and $T$ are the reflection and transmission coefficients, respectively, $A$ and $B$ represent the amplitudes of the pseudo-spinors in the nanomaterial with negative pseudospin parameter, $k_{x1} = +\sqrt{(E/\hbar v_F)^2 - \chi_1^2 k_y^2}$, $k_{x2} = +\sqrt{((E-\delta V)/\hbar v_F)^2 - \chi_2^2 k_y^2}$, $e^{i\theta_{q1}} = \hbar v_F (k_{x1} + i\chi_1 k_y)/|E|$ and $e^{i\theta_{q2}} = \hbar v_F (k_{x2} + i\chi_2 k_y)/|E - \delta V|$. We have assumed that the interfaces of nanomaterials are at $x = 0$ and $x = W$, and, for simplicity, that the energy is such that $0 < E < \delta V$. The angle of incidence $\theta_i$ can be determined from the electron velocity, $\mathbf{v} = \nabla_k E/\hbar$, and thus determines the transverse wave vector $k_y$ (e.g. for pristine graphene $k_y = (E/\hbar v_F)\sin\theta_i$). The unknown coefficients ($R$, $T$, $A$, and $B$) are determined by matching the pseudo-spinors at the interfaces: $\Psi|_{x=0^-} = \Psi|_{x=0^+}$ and $\Psi|_{x=L^-} = \Psi|_{x=L^+}$.

## *Figures*

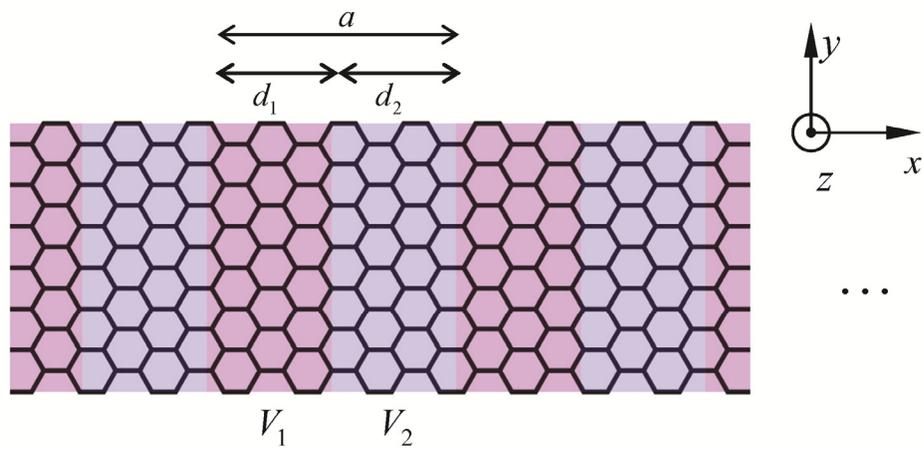

Fig. 1. Sketch of the geometry of a graphene superlattice characterized by a step-like periodic electrostatic potential.

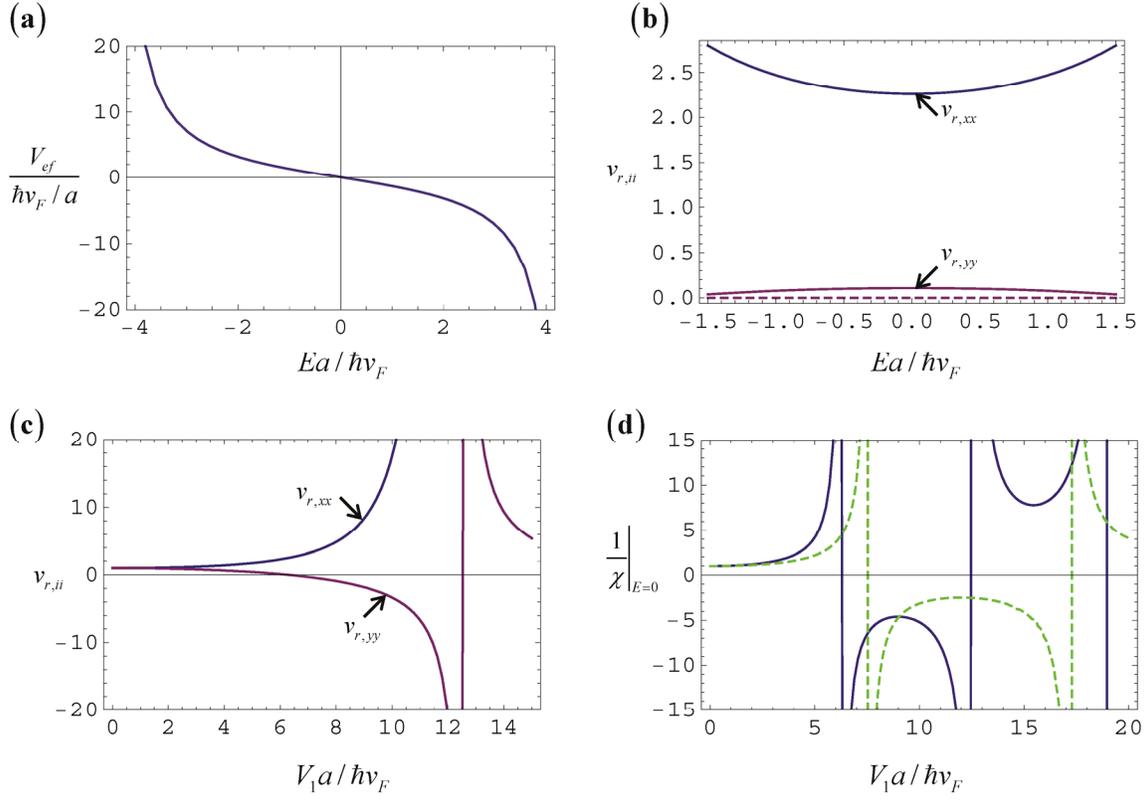

Fig. 2. (a) and (b): Effective parameters as function of the normalized energy ($E$) for a graphene superlattice with $d_1 = d_2$, $V_1 a / \hbar v_F = 6.0$, $V_2 = -V_1$. (a) Effective potential. (b) Parameters $v_{r,ii}$ (solid curves); the dashed curves represent the diagonal components of $\boldsymbol{\sigma}_{ef} \cdot \hat{\mathbf{x}}$ and $\boldsymbol{\sigma}_{ef} \cdot \hat{\mathbf{y}}$, which are very close to zero. (c) Parameters $v_{r,ii}$ as a function of $V_1$ for $E = 0$ and $V_1 = -V_2$ and $d_1 = d_2$. (d) Anisotropy ratio $1/\chi = v_{r,xx}/v_{r,yy}$ for ($i$) (solid curve) a Kronig-Penney electrostatic potential with $V_1 = -V_2$ and $d_1 = d_2$; ($ii$) (dashed curve) an electrostatic potential of the form $V = V_1 \sin(2\pi x / a)$.

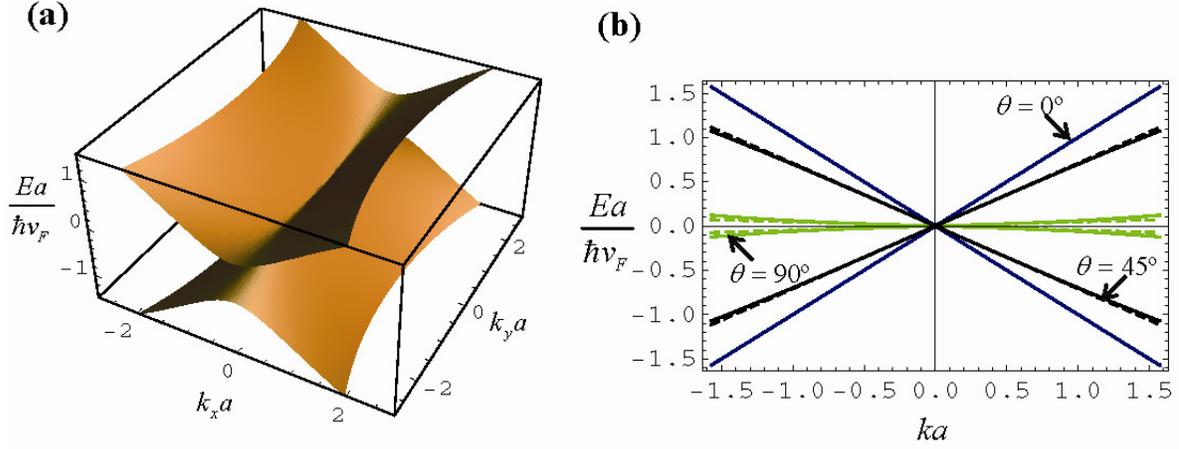

Fig. 3. (a) Exact energy dispersion of a graphene superlattice such that $d_1 = d_2$, $V_2 = -V_1$ and $V_1 a / \hbar v_F = 6.0$ (b) Dispersion of the energy eigenstates for $\mathbf{k} = k(\cos\theta, \sin\theta)$ and $V_1 a / \hbar v_F = 6.0$ calculated with (*i*) (solid curves) the "exact" energy dispersion characteristic of the superlattice. (*ii*) (dashed curves) the effective medium model based on the parameters $V_{ef}$ and $\boldsymbol{\sigma}_{ef}$.

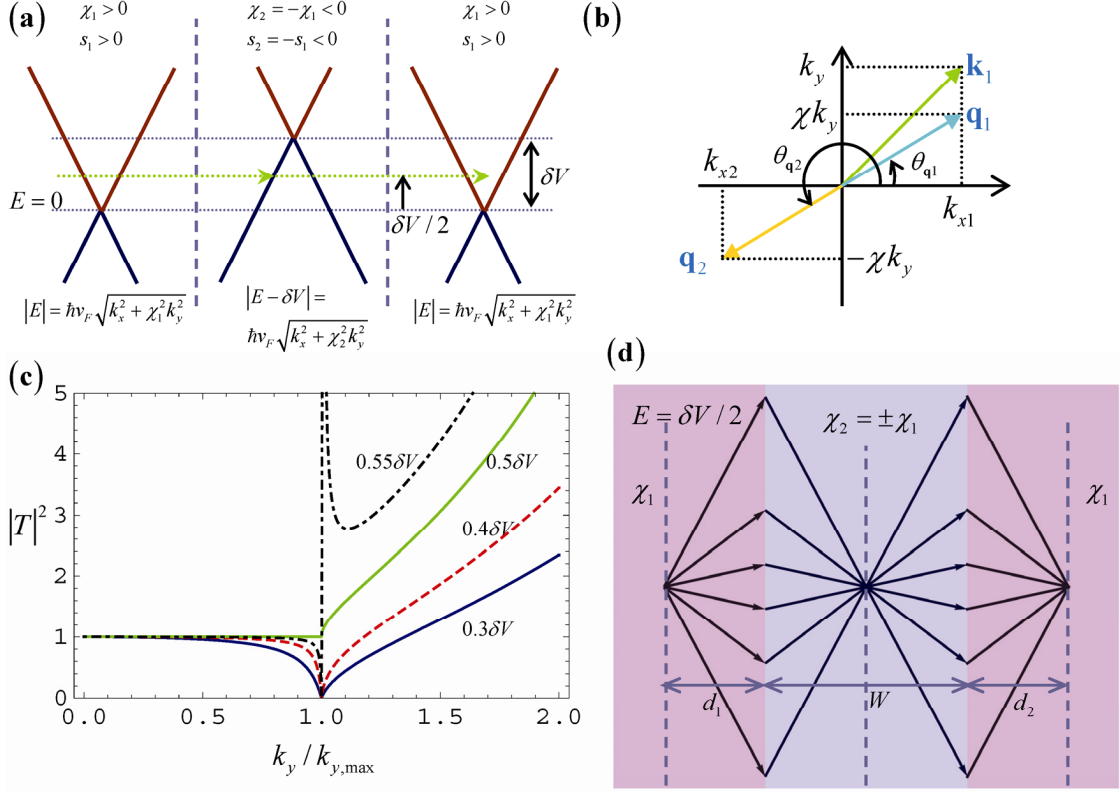

Fig. 4. Perfect tunneling through two complementary graphene nanomaterials. (a) Energy diagrams in the different regions. An electron with energy $E = \delta V/2$ can tunnel perfectly through the structure formed when two nanomaterials with symmetric anisotropy ratio ($\chi_1 = -\chi_2$) are paired. In material 1 the electron propagates in the conduction band ($s_1 > 0$), whereas in material 2 it propagates in the valence band ($s_2 < 0$). (b) Illustration of the property $\theta_{q2} = \theta_{q1} + \pi$ for $\chi \equiv \chi_1 = -\chi_2$ and $E = \delta V/2$. (c) Transmissivity as a function of the normalized wave vector component $k_y$ in material 1, for electrons with energy $E$ (travelling in material 1) that impinge on a slab of thickness $W$ of the material 2 (see panel (d)). The small text insets indicate the value of the normalized energy $E/\delta V$. It is assumed that $\chi_1 = -\chi_2 = 1/4.6$ and that the potential offset is such that $\delta V W/\hbar v_F = 1.0$. The normalization factor $k_{y,\max}$ is defined as the maximum value of the transverse momentum $k_y$: $k_{y,\max} = E/(\chi_1 \hbar v_F)$. For $k_y > k_{y,\max}$ the wave function decays exponentially and thus cannot be normalized (see the main text for a discussion). (d) Semi-classical picture of the electron trajectories in the nanostructure when $\chi_1 = |\chi_2|$ and $E = \delta V / 2$. The electrons are refracted at the interfaces with $\theta_t = -\theta_i$, analogous to Veselago-Pendry lens for photons.